\documentclass[useAMS,usenatbib,usegraphicx]{mn2e}

\newcommand{\morgana}{{\sc morgana}}

\newcommand{\be}{\begin{equation}}
\newcommand{\ee}{\end{equation}} 
\newcommand{\bea}{\begin{eqnarray}}
\newcommand{\eea}{\end{eqnarray}}

\def\lesssim{\,\lower2truept\hbox{${<\atop\hbox{\raise4truept\hbox{$\sim$}}}$}\,}
\def\gtrsim{\,\lower2truept\hbox{${>\atop\hbox{\raise4truept\hbox{$\sim$}}}$}\,}

\title[SFR function from GOODS-MUSIC]{On the Evolution of the SFR
  Function of Massive Galaxies. Constraints at $0.4<z<1.8$ from the
  GOODS-MUSIC Catalogue}

\author[Fontanot et al.]{
\parbox[t]{\textwidth}{ 
Fabio Fontanot$^{1,2}$, 
Stefano Cristiani$^1$,
Paola Santini$^3$,
Adriano Fontana$^3$,
Andrea Grazian$^3$ and
Rachel S. Somerville$^{4,5}$
}
\vspace*{6pt}\\
$^1$INAF-Osservatorio Astronomico, Via Tiepolo 11, I-34131 Trieste, Italy \\
$^2$HITS-Heidelberger Institut f\"ur Theoretische Studien, Schloss-Wolfsbrunnenweg 35, 69118 Heidelberg, Germany \\
$^3$INAF-Osservatorio Astronomico di Roma, via Frascati 33, I-00040, Monteporzio, Italy \\ 
$^4$Space Telescope Science Institute, 3700 San Martin Drive, Baltimore, MD 21218 \\
$^5$Department of Physics and Astronomy, Johns Hopkins University, Baltimore, MD 21218, USA \\
  email: fabio.fontanot@h-its.org}

\begin{document}
\date{Accepted ... Received ...}

\maketitle
\begin{abstract} 
We study the evolution of the Star Formation Rate Function (SFRF) of
massive ($M_\star>10^{10} M_\odot$) galaxies over the $0.4<z<1.8$
redshift range and its implications for our understanding of the
physical processes responsible for galaxy evolution. We use
multiwavelength observations included in the GOODS-MUSIC catalogue,
which provides a suitable coverage of the spectral region from 0.3 to
24 $\mu$m and either spectroscopic or photometric redshifts for each
object. Individual SFRs have been obtained by combining UV and 24
$\mu$m observations, when the latter were available. For all other
sources an ``SED fitting'' SFR estimate has been considered. We then
define a stellar mass limited sample, complete in the $M_\star>10^{10}
M_\odot$ range and determine the SFRF using the $1/V_{\rm max}$
algorithm. We thus define simulated galaxy catalogues based on the
predictions of three different state-of-the-art semi-analytical models
of galaxy formation and evolution, and compare them with the observed
SFRF. We show that the theoretical SFRFs are well described by a
double power law functional form and its redshift evolution is
approximated with high accuracy by a pure evolution of the typical SFR
(SFR$^\star$). We find good agreement between model predictions and
the high-SFR end of the SFRF, when the observational errors on the SFR
are taken into account. However, the observational SFRF is
characterised by a double peaked structure, which is absent in its
theoretical counterparts. At $z>1.0$ the observed SFRF shows a
relevant density evolution, which is not reproduced by SAMs, due to
the well known overprediction of intermediate mass galaxies at
$z\sim2$. Semi-analytical models are thus able to reproduce the most
intense SFR events observed in the GOODS-MUSIC sample and their
redshift distribution. At the same time, the agreement at the low-SFR
end is poor: all models overpredict the space density of SFR $\sim 1
M_\odot/yr$ and no model reproduces the double peaked shape of the
observational SFRF. If confirmed by deeper IR observations, this
discrepancy will provide a key constraint on theoretical modelling of
star formation and stellar feedback.
\end{abstract}


\section{Introduction}
The evolution of the star formation rate (SFR) over the cosmic time is
a fundamental constraint for every theory of galaxy formation and
evolution (see e.g. \citealt{Hopkins04, HopkinsBeacom06}). Estimating
SFRs for individual galaxies is a complex task, owing to the
uncertainties involved in the reconstruction of this quantity from
observational data. It is widely accepted that dusty molecular clouds
are the main sites for star formation: this implies that newly born
stars are subject to significant dust attenuation, until they are able
to escape or disrupt their parent cloud. Young OB stars emit a
considerable amount of energy in the restframe ultraviolet (UV) band,
which has thus been considered a key waveband for recovering the
instantaneous SFR in external galaxies. However, dust absorbs UV
photons, heats up and re-emits this energy as thermal radiation in the
Infrared (IR) bands. For this reason, SFR estimates based on UV
luminosity include correction factors to account for dust attenuation
and re-emission \citep{Calzetti94, Kennicutt98,Bell03}. Dust
attenuation is particularly relevant for star forming galaxies, where
the dust emission peak is the dominant component of the galactic
spectral energy distribution (SED, see e.g. \citealt{Calzetti00}), not
to mention the extreme cases of sub-mm galaxies (see
e.g. \citealt{Chapman04}). The study of the cosmic IR background shows
that the global energy emitted by galaxies in the IR is comparable to
the direct starlight emission, detectable in the optical (see
e.g. \citealt{Lagache05}), clearly showing that a significant fraction
of star formation activity is expected to be heavily extinguished and
detectable only in the IR.

These uncertainties become more relevant, when SFR estimates for
galaxies covering a wide redshift range are considered, since both
galaxy physical properties and dust properties are expected to evolve
with cosmic time \citep{Maiolino04, Fontanot09a, Gallerani10,
  Fontanot11c}. Combining UV information with supplementary
information from direct observations in the IR region has thus been
proposed as the best tool to account for the total SFR
\citep{Kennicutt98,Bell07}. In particular, observations at $24 \mu$m
have shown to be extremely useful for estimating the global IR
luminosity and the instantaneous SFR \citep{Papovich07,Santini09}.
Thanks to the advent of the {\it Herschel} Space Observatory, we will
be able to constrain total IR luminosities, by directly sampling the
peak of the thermal emission. In fact, it has been recently shown by
\citet{Rodighiero10b}, that the combination of $24 \mu$m data with
      {\it Herschel} observations at longer wavelengths represents a
      very promising tool for a more accurate determination of SFRs of
      individual galaxies up to $z\sim3$.

Nowadays, our view of the evolution of galaxy properties has
substantially changed thanks to the advent of multiwavelength
surveys. If the spectral sampling is fine enough, these catalogues can
be used to infer the physical properties also in the absence of
spectroscopic information. In recent years several groups (see
e.g. \citealt{Fontana04,Panter07}) developed a number of ``SED
fitting'' algorithms: these codes compare the available photometry for
individual sources with synthetic SED libraries and the best-fit model
template is chosen by means of a $\chi^2$ minimisation. The resulting
estimate for redshift (the so-called photometric redshift), stellar
mass ($M_\star$), dust extinction and SFR are the most widely used
results of this procedure and have been of fundamental importance for
our understanding of galaxy evolution. 

The common interpretation of these results is connected to the
so-called ``downsizing'' scenario, in which the star formation shifts
from high mass to low mass galaxies as redshift decreases (first
introduced by \citealt{Cowie96}). This picture has been recently
revised by \citet{Fontanot09b} by showing that the typical errors
associated with the estimate of the physical quantities has to be
taken into account, when comparing the observed evolutionary trends
with the predictions of theoretical models. In particular,
\citet{Fontanot09b} concluded that the discrepancies seen between
models and data for massive galaxies are not significant, if model
predictions are convolved with typical observational errors, while the
strongest discrepancies between model and data are seen for
low-to-intermediate mass galaxies. 

The importance of the study of the redshift evolution of the cosmic
SFR for galaxies at a given mass range has been widely recognised by a
number of authors \citep{Noeske07, Elbaz07, Zheng07, Drory08, Dunne08,
  Santini09, Gilbank11}. These results and the evolution of the
stellar mass density provide fundamental information about the
evolution of the global process of galaxy formation. The latter
approach is best complemented by the analysis of the stellar mass
function (i.e. the volume density of galaxies as a function of stellar
mass), which allows us to characterise galaxy evolution as a function
of both redshift and stellar mass. Despite the wealth of information
already available in the literature about the stellar mass function
and its redshift evolution, very little is known about the
corresponding Star Formation Rate Function (SFRF hereafter) and its
evolution: \citet{Bell07} studied its evolution in the $0.2<z<1.0$
redshift range in the COMBO-17 survey, \citet{Schiminovich07}
presented the distribution of SFRs in the SDSS survey and
\citet{Bothwell11} considered the distribution of SFRs in the Local
Volume (defined as an 11 Mpc radius around the Milky Way) combining
GALEX and {\it Spitzer} observations. This is mainly due to the
uncertainties involved in the SFR estimate, but also to the
difficulties in correctly accounting for the completeness correction
needed, especially at the low-SFR end.

In this paper, we focus on a new treatment of the GOODS-MUSIC data, by
focusing on a mass complete sample of galaxies. In this way, we are
able to fully characterise the various levels of star formation in
these objects and follow their redshift evolution. Moreover, since the
strongest starbursts are hosted by galaxies above the chosen mass
limit, the high-SFR end of the resulting SFRF provides fundamental
information about the redshift evolution of this particular galaxy
class. Therefore, this paper represents a complementary study of the
evolution of SFRs in galaxies of different stellar mass with respect
to \citet{Santini09}, to which we refer the reader for the analysis of
the evolution of the cosmic SFR and specific SFR. A detailed
comparison of the same catalogue with other estimates for the
evolution of SFR as a function of redshift and stellar mass has been
presented in \citet{Fontanot09b} (see references herein). In the
present paper we focus on the determination of the SFRF and compare
results with the predictions of state-of-the-art semi-analytical
models of galaxy formation and evolution.

This paper is organised as follows. We present the GOODS-MUSIC
catalogue in sec.~\ref{sec:dataset} and the theoretical predictions in
sec.~\ref{sec:models}; we present our results in
sec.~\ref{sec:results}; finally we discuss our conclusions in
sec.~\ref{sec:summary}. In the original GOODS-MUSIC papers a Salpeter
initial mass function (IMF) is mostly used; throughout this work, we
assume a Chabrier IMF and we rescale all stellar mass and SFR
estimates, accordingly.  We also adopt the $\Lambda$CDM concordance
cosmological model ($H_0 = 70 \, km/s/Mpc$, $\Omega_m=0.3$ and
$\Omega_\lambda=0.7$).

\section{Dataset}\label{sec:dataset}

Our dataset is based on the updated version of GOOD-MUSIC catalogue
\citep{Grazian06,Santini09}. GOODS-MUSIC is a multiwavelength
catalogue, with photometry in 15 bands, from 0.35 to 24 $\mu$m,
covering an area of $\sim 143.2 arcmin^2$ of the GOODS-South field and
listing $\sim$ 15000 sources. Main source detection has been performed
on the z image, but faint z-band objects detected either on the K
image or at 4.5 $\mu$m have also been included in the catalogue. The
GOODS-MUSIC catalogue has been also cross-correlated with available
spectroscopic catalogues, and a spectroscopic redshift has been
assigned to $\sim 12\%$ of sources. For each source in the catalogue,
its physical parameters, such as $M_\star$, SFR and the photometric
redshift, have been estimated through a ``SED fitting'' algorithm (see
e.g. \citealt{Santini09} and reference herein); for the purposes of
$\chi^2$ minimisation, the available broad band photometry up to
rest-frame $\lambda<5.5 \mu$m has been considered. The reference
synthetic SED library is defined using \citet{Bruzual03} synthetic
models and assuming exponentially declining star formation
histories. The relevant parameters for the SED synthesis (such as dust
content, metalicity, age, timescale for star formation histories, etc)
are defined on a grid in order to cover a wide range of possible
values and combinations. The discreteness of SED generation implies
that the final estimates for the physical parameters are subject to
relevant degeneracies (see e.g. \citealt{Marchesini09} for a
discussion focused on the effect of changing the physical ingredients
in the adopted stellar population models). In particular, theoretical
models predict star formation histories for individual galaxies, which
grossly differ from any assumed smooth analytical shape: this has
important implications for the $M_\star$ estimate
\citep{Stringer09}. By comparing the available spectroscopic redshifts
($z_{\rm spec}$) with the corresponding estimates from the SED fitting
algorithm ($z_{\rm phot}$), \citet{Santini09} computed the average
absolute scatter between spectroscopic and photometric redshift
$\Delta z = \frac{|z_{\rm spec}-z_{\rm phot}|}{1+z_{\rm spec}}$. They
found $ \sigma(\Delta z) = 0.06$ for the whole sample and
$\sigma(\Delta z) = 0.043$ when only the brightest galaxies ($z<23.5$)
are considered.
 
A parallel estimate of the star formation rate is provided by the
combination of IR and UV luminosities. For sources in the GOODS-MUSIC
sample, \citet{Santini09} combined the rest-frame UV luminosity
derived from the SED fitting and the integrated luminosity between 8
and 1000 micron, obtained by fitting 24 micron fluxes with
\citet{Dale02} templates. The correction provided by
\citet{Papovich07} has been applied on the resulting SFRs in order to
account for the overestimation of the total IR luminosity at high
redshifts/IR luminosities when extrapolating from 24 $\mu$m fluxes
(e.g. \citealt{Elbaz10}). \citet{Santini09} explicitly compared these
SFR estimates with the SED fitting results, for sources with both
estimates available. They found an overall good agreement, but also a
systematic trend at all redshift, with IR-based estimates exceeding
the fit-based ones as the star formation rate increases, and vice
versa for low values of SFR (see also \citealt{Papovich07,
  Nordon10}). In the following of this paper, unless explicitly
stated, for each object in the sample we consider IR-based SFRs when
available, and ``SED fitting'' estimates elsewhere.

In order to provide a reliable estimate of statistical quantities like
the stellar mass function and the SFR function, a careful analysis of
observational biases and selection effects is of fundamental
importance for a complete census of all galaxies. Those corrections
are critical for the SFRF, given the complex combination of
multiwavelength information needed, including both the selection of
galaxies in the $K$-band, fundamental for the definition of a
mass-limited sample, and the SFR estimate from $24 \mu$m for direct
detection (if available) or UV and optical data for the SED fitting
procedure. In order to reduce the uncertainties involved in our
analysis, we decide to focus on a subsample of the GOOD-MUSIC
catalogue, defined by considering only sources with a stellar mass
estimate $M_{\star} > 10^{10} M_\odot$. The GOODS-MUSIC catalogue is
expected to be highly complete in this mass range over the whole
$0.4<z<1.8$ redshift interval, corresponding to $K$-band magnitudes
brighter than $\sim$23 \citep{Fontana06}. Moreover, for these objects,
either $24 \mu$m fluxes or detailed multiwavelength photometry are
available, and provide reliable estimates for SFR$> 0.01 M_\odot/yr$
\citep{Santini09}. Objects with documented AGN activity
(i.e. spectroscopically determined or X-ray sources) have been removed
from the final sample. Compton thick candidates, selected on the basis
of their very red spectra \citep{Fiore08}, have been removed as
well. In the redshift range of our interest there are only 4 removed
objects with $1.4<z<1.8$ ($0.5\%$ of the considered sample). We test
that our conclusions are robust against the inclusion of these
objects. Our final sample consists of 740 sources.

It is also worth mentioning the known overdensities at $z \sim 0.7$
and $z \sim 1.6$ in the GOODS fields, as well as the known
underdensity at $z \sim 0.9$ \citep{Gilli03, Castellano07,
  Salimbeni09}. Ongoing surveys like COSMOS \citep{Scoville07} and
CANDELS \citep{Grogin11,Koekemoer11}, will be able to reduce the
effect of cosmic variance on the SFRF determination, and study its
environmental dependencies.

\section{Models}\label{sec:models}
\begin{figure}
  \centering
  \includegraphics[width=7.25cm]{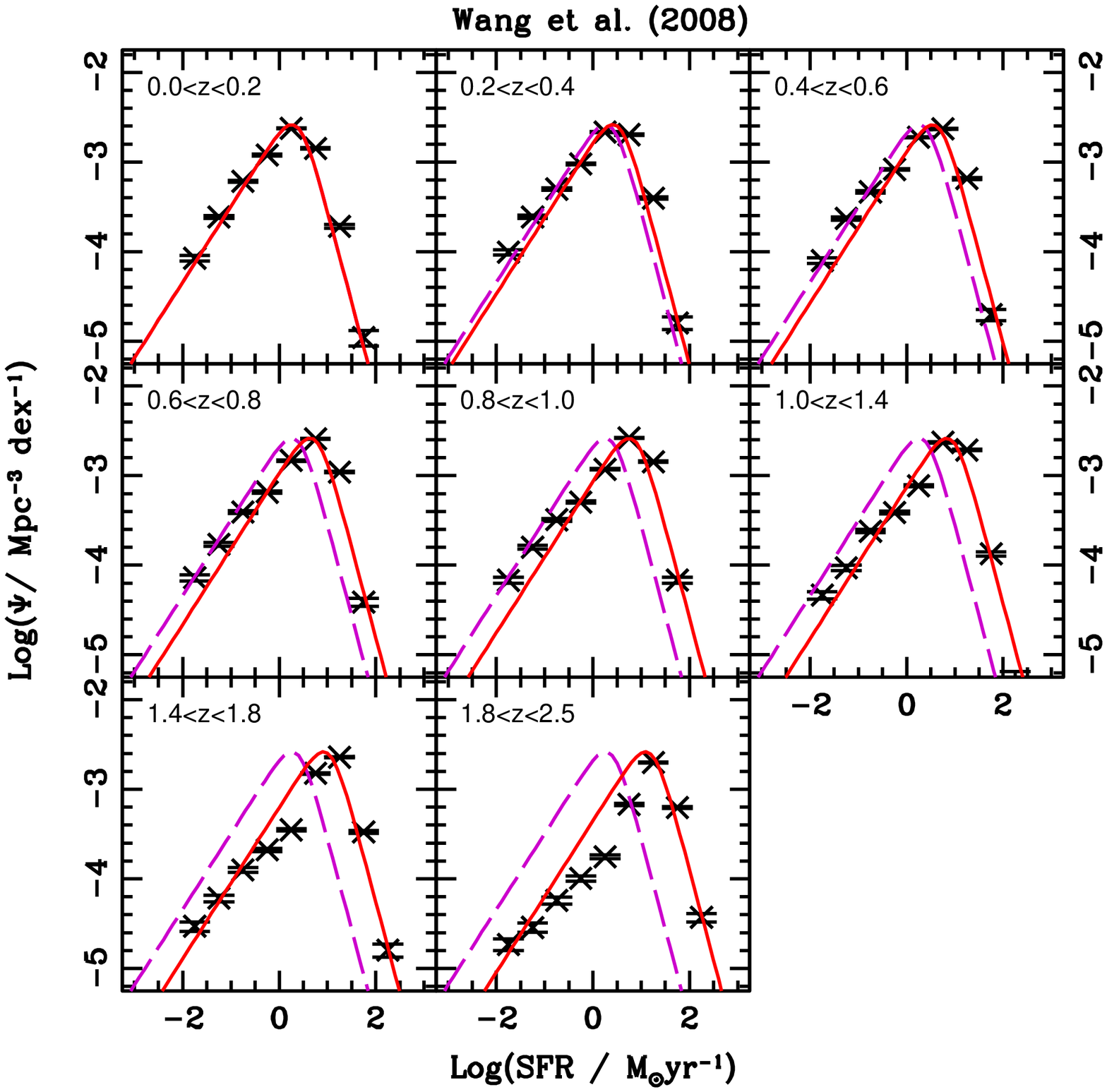}
  \includegraphics[width=7.25cm]{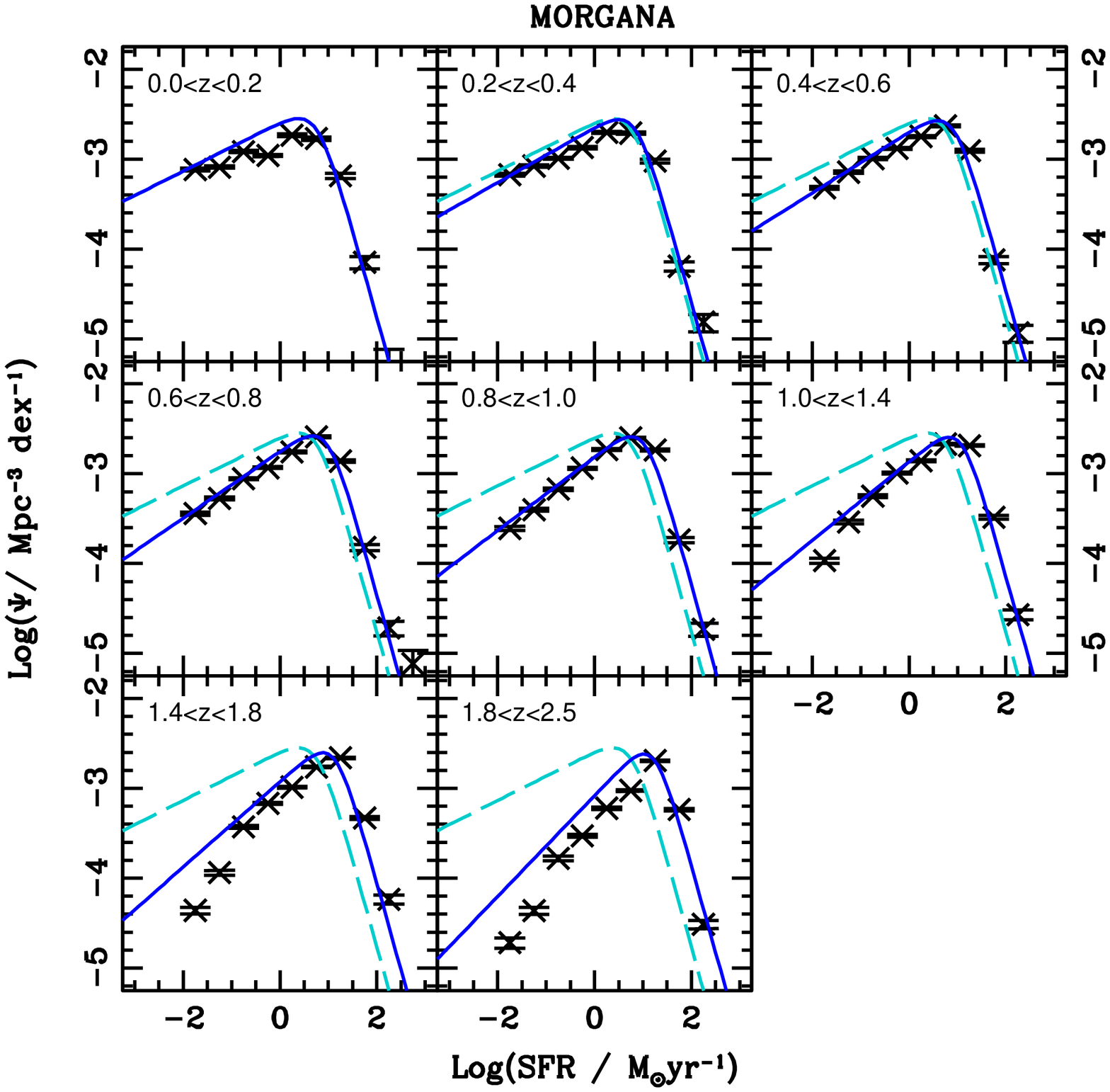}
  \includegraphics[width=7.25cm]{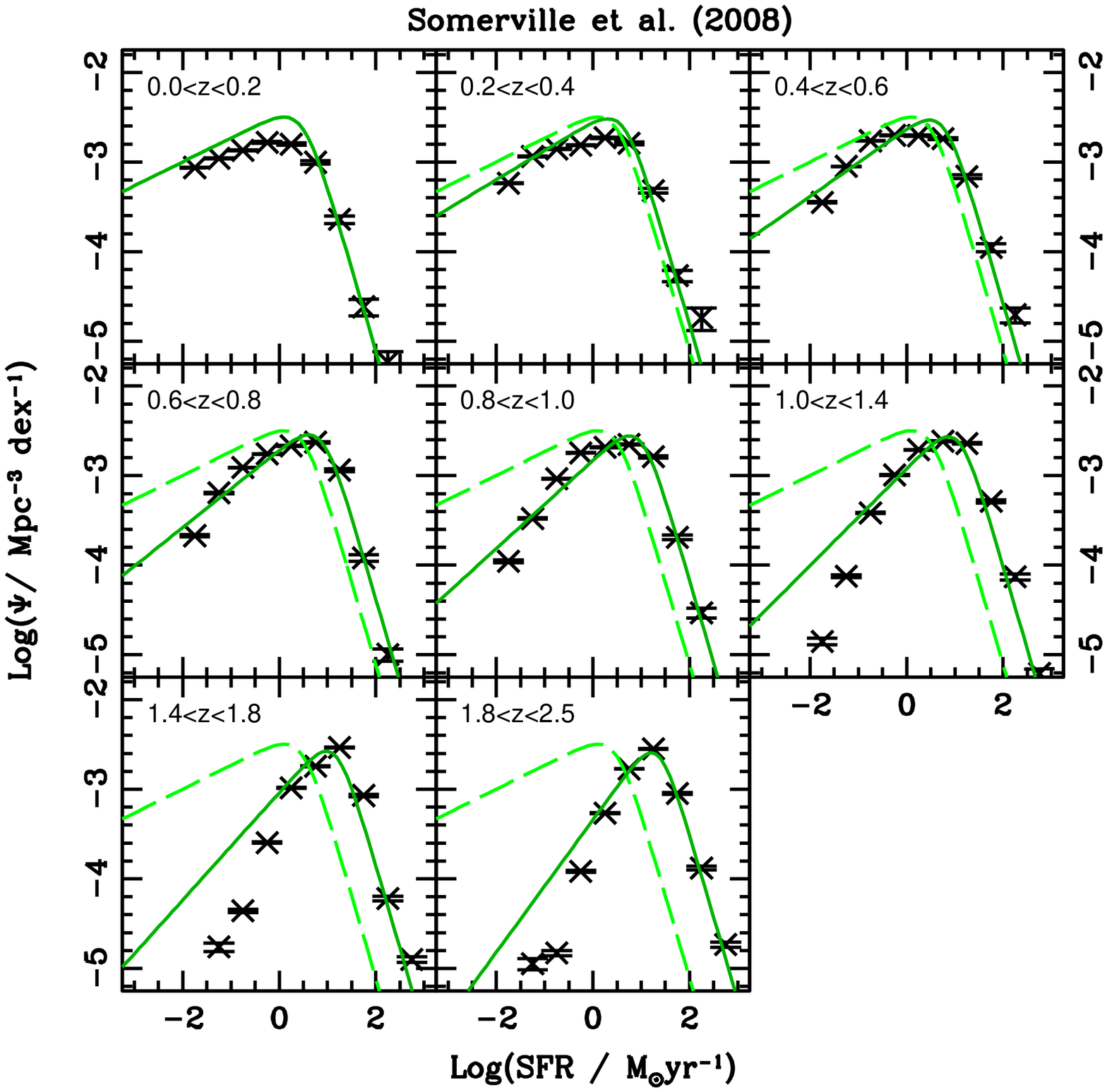}
\caption{The redshift evolution of the SFRF in semi-analytical
  models. Crosses refer to the intrinsic predictions of the models,
  while solid lines show the analytic fit (pure SFR$^\star$ evolution)
  of the theoretical SFRF. In each panel, dashed line reports the
  position of the $z=0$ SFRF.
  \label{fig:theo_sfrf}}
\end{figure}

In order to understand the physical implications of our results, we
consider the prediction of the recent implementations of 3
state-of-the-art semi-analytical models, namely, the \citet{Wang08}
model, the \citet{Somerville08} and \morgana~\citep{Monaco07}. The
predictions of these 3 models have been extensively compared in a
number of recent papers \citep{Kimm08, Fontanot09b, Fontanot10} and we
refer the reader to these publications and the original papers for all
the details of the galaxy formation and evolution modelling. In these
models, the evolution of the baryonic component is followed by means
of an approximate description of the physical processes at play (such
as gas cooling, star formation, stellar feedback, Black Hole growth
and AGN feedback) and of their interplay with gravitational processes
(i.e. dynamical friction, tidal stripping and two-body mergers),
linked to the assembly of the large-scale structure of the
Universe. These ``recipes'' include a number of parameters which are
usually fixed by comparing model predictions with a set of
low-redshift observations. Despite their simplified approach, SAMs
have turned into a flexible and widely used tool to explore a broad
range of specific physical assumptions, as well as the interplay
between different physical processes. All models we consider resolve
galaxies with $M_\star > 10^{9} M_\odot$, and they reproduce fairly
well a number of observational constraints on the evolution of the
SFR, both as a global quantity (i.e. the cosmic star formation rate),
and for individual galactic population (i.e. the fraction of passive
galaxies in the local Universe, see e.g. \citealt{Kimm08}).

Also for the theoretical predictions, we consider purely stellar mass
selected samples ($M_\star > 10^{10} M_\odot$) and, as for the
GOOD-MUSIC catalogue, we excluded from the analysis all objects with
SFR$<0.01 M_\odot/yr$. After the selection procedure we are thus left
with 385187 galaxies for the \citet{Wang08} model, 342795 for the
\citet{Somerville08} model and 249823 galaxies for \morgana. In
fig.~\ref{fig:theo_sfrf}, we show the binned SFRFs for each model
(crosses). In order to quantitatively estimate the redshift evolution
of the predicted SFRF we apply to model predictions a fitting
procedure similar to \citet{Fontanot07a}. We determined best fit
analytical solutions to the theoretical predictions assuming both a
Schechter and a double power law shape for the SFRF. We use the
corresponding number densities to Monte Carlo generate independent
data points in the z-SFR space and we calibrate the best fit
parameters for the SFRF and its evolution by means of a $\chi^2$
binning procedure. Our results show that the double power law shape
systematically scored better results than the Schechter function in
terms of $\chi^2$ (with typical values $\chi^2<1.0$). This is mainly
due to the extra degree of freedom in the high-SFR-end slope, with
provides a better description of this region with respect to the
exponential cut-off of the Schechter function, even if it is worth
stressing that the Schechter shape is not formally ruled out by our
analysis. Anyhow, in the following we thus assume a SFRF in the form
of a double power-law:
\begin{equation}
\Phi(SFR) =
\frac{\Phi^\star(SFR^\star)}{(SFR/SFR^\star)^{-\alpha}+(SFR/SFR^\star)^{-\beta}}
\end{equation}
where $\Phi^\star$, SFR$^\star$, $\alpha$, $\beta$ are free
parameters. In order to model the redshift evolution of the SFRF we
fix the values of the 4 parameters at their $z=0$ values and we
consider different combinations for their redshift evolution. We
consider both a pure evolution in SFR$^\star$ (PSE) of the form:
\begin{equation}
SFR^\star = SFR^\star_{z=0} (1+z)^{k_{\rm SFR}}
\end{equation}
and a pure density evolution (PDE) of the form:
\begin{equation}
\Phi^\star = \Phi^\star_{z=0} e^{k_{\Phi}(1+z)}
\end{equation}
where $k_{\rm SFR}$ and $k_{\Phi}$ represents the evolutionary
parameters. Our results point out that the redshift evolution of the
theoretical SFRF is in general well described by a pure evolution in
SFR$^\star$.  In fig.~\ref{fig:theo_sfrf}, we show the best fit
results for the \citet{Wang08} model as a solid red line, while the
dashed magenta line represents the corresponding $z=0.1$ best fit
SFRF. As for \morgana~and the \citet{Somerville08} model, the fit at
the low-SFR end is improved (by a factor of 2 in terms of $\chi^2$) if
the pure SFR$^\star$ evolution (which is still statistically
acceptable) is combined with an evolution of the slope $\alpha$ of the
form:
\begin{table*}
  \caption{Best fit parameters for the evolution of the SFR function
    as predicted by semi-analytical models.}
  \label{tab:bf_theo}
  \renewcommand{\footnoterule}{}
  \centering
  \begin{tabular}{ccccccccc}
    \hline
    SAM & $\Phi^{\star} (\times 10^{-2})$ & $SFR^{\star}$ & $k_{\rm SFR}$ & $\alpha$ & $\beta$ & $\alpha_0$ & $k_\alpha$ & $\chi^2$\\
    \hline
    \citet{Wang08}       & $0.47 \pm 0.04$ & $0.29 \pm 0.14$ & $2.10 \pm 0.64$ & $0.84 \pm 0.10$ & $-2.03 \pm 0.27$ & --- & --- & $0.32$\\
    \citet{Somerville08} & $0.46 \pm 0.05$ & $0.39 \pm 0.12$ & $1.99 \pm 0.72$ & --- & $-1.84 \pm 0.24$ & $0.29 \pm 0.07$ & $0.23 \pm 0.11$ & $0.44$\\
    \morgana             & $0.41 \pm 0.03$ & $0.70 \pm 0.17$ & $1.23 \pm 0.42$ & --- & $-1.95 \pm 0.22$ & $0.26 \pm 0.03$ & $0.15 \pm 0.04$ & $0.29$ \\
    \hline
  \end{tabular}
\end{table*}
\begin{equation}
\alpha = \alpha_0 + k_\alpha z
\end{equation}
where $\alpha_0 = \alpha(z=0)$ and $k_\alpha$ is the evolutionary
parameter. We conclude that a double power law is a good
representation of the SFRF as predicted by the theoretical models at
$z<2$ and for SFR$>0.01 M_\odot/yr$. We present the best fit
parameters corresponding to the above equations in
tab.~\ref{tab:bf_theo}. It is worth stressing that we obtain very low
$\chi^2$ value even if our fits tend to break down at the highest
redshift and lowest SFRs. This is due to the high number of object in
the model samples, which thus provide a very accurate fit of the knee
of the SFRF and its evolution.

In order to compare the theoretical predictions with the results from
the GOODS-MUSIC catalogue, we repeat the fitting procedure after
convolving the theoretical predictions with an estimate of the
systematic error on the SFR determination, to fully account for the
typical observational error in the catalogue \citep{Santini09}. We
assume a log-normal error distribution for the SFR with an amplitude
of 0.3 dex (see also \citealt{Fontanot09b}). We repeat the above
analysis and we determine that a pure SFR$^\star$ evolution of a
double power law is still a good description (i.e. consistent $\chi^2$
values) of the overall redshift evolution of the predicted SFRFs. The
most noticeable difference in the best fit parameters is the
systematic increase of the $\beta$ parameter by as much as $20\%$.

\section{Results \& Discussion}\label{sec:results}
\begin{figure*}
  \centering
  \includegraphics[width=15cm]{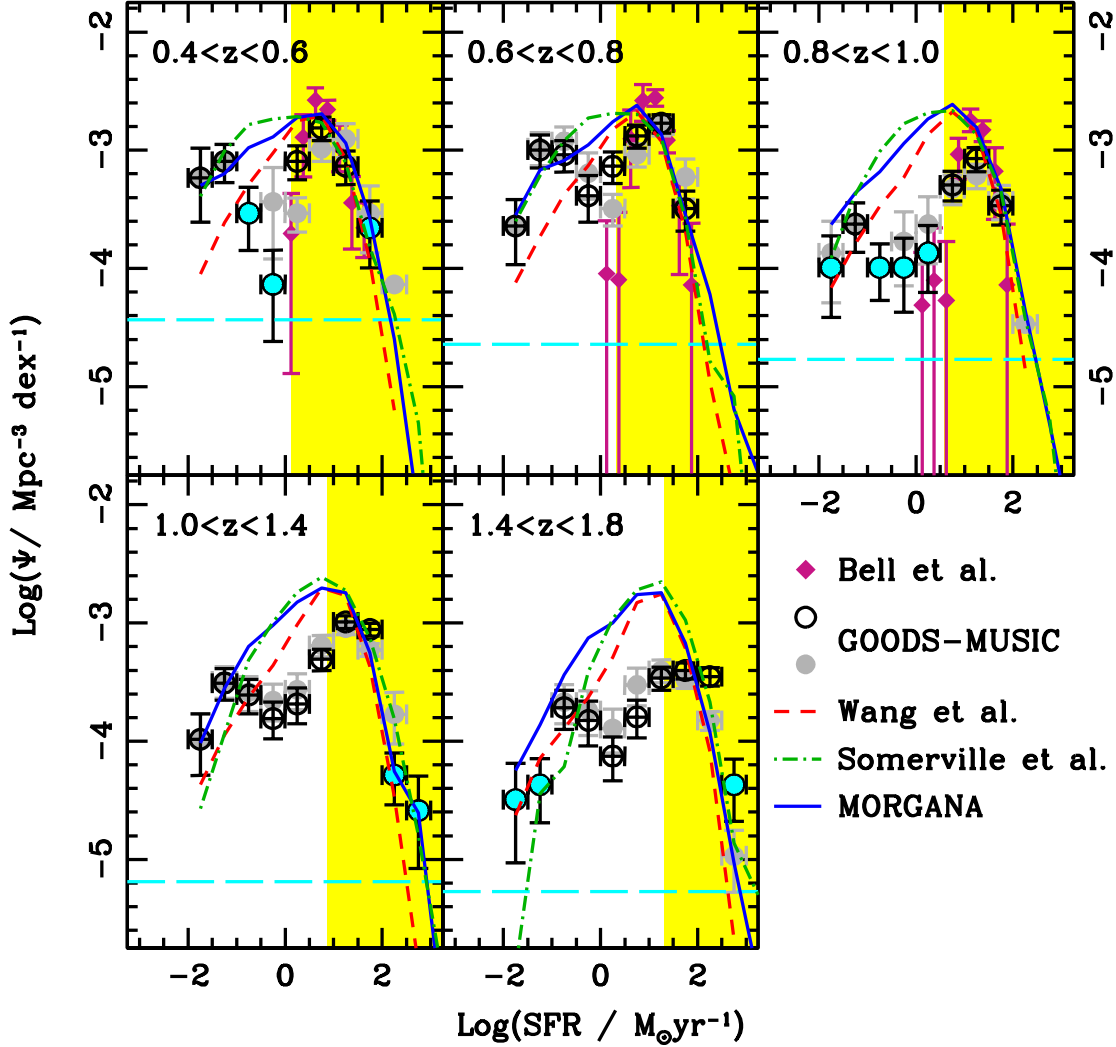}
\caption{The evolution of the SFRF for $M>10^{10} M_\odot$ galaxies in
  the GOOD-MUSIC sample. In each panel, the horizontal cyan
  long-dashed line represents the space density corresponding to one
  object in the GOODS-MUSIC volume, while the yellow shading
  highlights the the SFR range covered by $24 \mu$m observations in
  each redshift bin. Open circles refer to our SFRF estimate from the
  GOODS-MUSIC survey: in each subpanel, cyan solid circles mark the
  bins with less than 5 galaxies; errorbars represent the 1-$\sigma$
  confidence levels (see text for more details). Light grey filled
  circles refer to the SFRF derived only from the results of the SED
  fitting algorithm. Purple diamonds refer to the SFRF estimate from
  the COMBO-17 \citep{Bell07} survey. In all panels, dashed red,
  dot-dashed green and solid blue lines refer to the \citet{Wang08},
  \citet{Somerville08} and \morgana~predictions, respectively. SAM
  predictions are convolved with an Gaussian error on the SFR (width
  $0.3$ dex, see text).
  \label{fig:sfrf}}
\end{figure*}
\begin{table}
  \caption{SFR Function in the GOODS-MUSIC Catalogue.}
  \label{tab:results}
  \renewcommand{\footnoterule}{}
  \centering
  \begin{tabular}{ccc}
    \hline
    & $log(SFR)$ & $Log(\Phi(SFR))$ \\
    \hline
    $0.4<z<0.6$ & $-1.75 \pm 0.25$ & $-3.23^{+0.25}_{-0.38}$ \\
                & $-1.25 \pm 0.25$ & $-3.09^{+0.14}_{-0.18}$ \\
                & $-0.75 \pm 0.25$ & $-3.53^{+0.22}_{-0.31}$ \\
                & $-0.25 \pm 0.25$ & $-4.14^{+0.29}_{-0.48}$ \\
                & $+0.25 \pm 0.25$ & $-3.09^{+0.13}_{-0.16}$ \\
                & $+0.75 \pm 0.25$ & $-2.81^{+0.09}_{-0.10}$ \\
                & $+1.25 \pm 0.25$ & $-3.14^{+0.13}_{-0.16}$ \\
                & $+1.75 \pm 0.25$ & $-3.69^{+0.23}_{-0.34}$ \\
    \hline
    $0.6<z<0.8$ & $-1.75 \pm 0.25$ & $-3.64^{+0.23}_{-0.33}$ \\
                & $-1.25 \pm 0.25$ & $-2.99^{+0.13}_{-0.15}$ \\
                & $-0.75 \pm 0.25$ & $-3.04^{+0.12}_{-0.15}$ \\
                & $-0.25 \pm 0.25$ & $-3.39^{+0.17}_{-0.22}$ \\
                & $+0.25 \pm 0.25$ & $-3.14^{+0.12}_{-0.15}$ \\
                & $+0.75 \pm 0.25$ & $-2.88^{+0.09}_{-0.10}$ \\
                & $+1.25 \pm 0.25$ & $-2.77^{+0.07}_{-0.08}$ \\
                & $+1.75 \pm 0.25$ & $-3.50^{+0.15}_{-0.19}$ \\
    \hline
    $0.8<z<1.0$ & $-1.75 \pm 0.25$ & $-3.99^{+0.27}_{-0.42}$ \\
                & $-1.25 \pm 0.25$ & $-3.63^{+0.18}_{-0.24}$ \\
                & $-0.75 \pm 0.25$ & $-3.99^{+0.20}_{-0.28}$ \\
                & $-0.25 \pm 0.25$ & $-3.99^{+0.25}_{-0.38}$ \\
                & $+0.25 \pm 0.25$ & $-3.87^{+0.23}_{-0.34}$ \\
                & $+0.75 \pm 0.25$ & $-3.30^{+0.12}_{-0.14}$ \\
                & $+1.25 \pm 0.25$ & $-3.07^{+0.10}_{-0.11}$ \\
                & $+1.75 \pm 0.25$ & $-3.47^{+0.13}_{-0.16}$ \\
    \hline
    $1.0<z<1.2$ & $-1.75 \pm 0.25$ & $-3.99^{+0.21}_{-0.30}$ \\
                & $-1.25 \pm 0.25$ & $-3.51^{+0.12}_{-0.15}$ \\
                & $-0.75 \pm 0.25$ & $-3.61^{+0.13}_{-0.16}$ \\
                & $-0.25 \pm 0.25$ & $-3.81^{+0.14}_{-0.17}$ \\
                & $+0.25 \pm 0.25$ & $-3.68^{+0.14}_{-0.17}$ \\
                & $+0.75 \pm 0.25$ & $-3.31^{+0.08}_{-0.10}$ \\
                & $+1.25 \pm 0.25$ & $-2.99^{+0.05}_{-0.06}$ \\
                & $+1.75 \pm 0.25$ & $-3.06^{+0.05}_{-0.06}$ \\
                & $+2.25 \pm 0.25$ & $-4.29^{+0.19}_{-0.25}$ \\
                & $+2.75 \pm 0.25$ & $-4.59^{+0.29}_{-0.49}$ \\
    \hline
    $1.2<z<1.8$ & $-1.75 \pm 0.25$ & $-4.49^{+0.31}_{-0.54}$ \\
                & $-1.25 \pm 0.25$ & $-4.37^{+0.22}_{-0.32}$ \\
                & $-0.75 \pm 0.25$ & $-3.72^{+0.14}_{-0.18}$ \\
                & $-0.25 \pm 0.25$ & $-3.83^{+0.17}_{-0.21}$ \\
                & $+0.25 \pm 0.25$ & $-4.13^{+0.16}_{-0.21}$ \\
                & $+0.75 \pm 0.25$ & $-3.80^{+0.14}_{-0.18}$ \\
                & $+1.25 \pm 0.25$ & $-3.47^{+0.09}_{-0.10}$ \\
                & $+1.75 \pm 0.25$ & $-3.40^{+0.07}_{-0.07}$ \\
                & $+2.25 \pm 0.25$ & $-3.45^{+0.07}_{-0.08}$ \\
                & $+2.75 \pm 0.25$ & $-4.37^{+0.22}_{-0.31}$ \\
    \hline
  \end{tabular}
\end{table}

We first compute the binned SFRF for massive galaxies by means of the
standard $1/V_{\rm max}$ formalism, and we show the resulting SFRF in
fig.~\ref{fig:sfrf} as open circles (cyan circles mark the bin with
less than 5 sources). In order to take into account the observational
uncertainties, we repeat the analysis 10000 times after convolving
every individual SFR estimate with the corresponding error on SFR and
redshift (see \citealt{Santini09} for more details on the error
estimate). Errorbars in fig.~\ref{fig:sfrf} refer to the 1-$\sigma$
variance over these 10000 realisations. We collect our final estimate
of the SFRF and associated errors in tab.\ref{tab:results}. In the
fig.~\ref{fig:sfrf} we also include the previous estimate for the SFRF
at $z>0$ from the COMBO-17 survey \citep{Bell07}. It is worth noting
that the COMBO-17 results are obtained by considering all galaxies in
their sample, with no cut in stellar mass. Our estimates for the SFRF
are in good agreement with the \citet{Bell07} result at the high-SFR
end, where we expect the massive galaxies to dominate the space
density.

Fig.~\ref{fig:sfrf} clearly show that the COMBO-17 data are not deep
enough to explore the low-SFR end with the same extent as in the
GOOD-MUSIC catalogue, moreover the low-SFR end of the SFRF shows a
very peculiar feature, i.e. a double peaked shape. It is really
tempting to interpret this feature in terms of the well known
bimodality in the colour (SFR)-mass diagram, and/or connect those
galaxies with the intermediate sources populating the so-called {\it
  green valley}. It is also worth noting that a populations of
relatively massive galaxies with very low levels of star formation has
been already reported at lower redshifts \citep{Wilman08,
  Schawinski10}.

However, a word of caution is necessary. We have limited information
about individual $24 \mu$m fluxes for galaxies in our mass-selected
sample, and typically only for SFR$>1 M_\odot/yr$. This implies that
the secondary peak at SFR$<1 M_\odot/yr$ is populated by galaxies
whose SFR estimate comes from the SED fitting algorithm. In order to
test the robustness of our result, we repeat our analysis considering
only SFR estimates coming from the SED fitting. We find no significant
change in the shape of the resulting SFRF (light grey filled circles
in fig.~\ref{fig:sfrf}): in particular, the double peaked behaviour
persists. Our main conclusions are therefore insensitive to the origin
of SFR estimate, in agreement with \citet{Santini09} results (i.e. the
$24 \mu$m-based SFR estimates are in reasonable agreement with SED
fitting results for sources with both estimates available).

Nonetheless, we cannot completely exclude the hypothesis that the gap
between the two peaks is an artifact of the SED fitting procedure
(i.e. piling up of low-SFR galaxies in the secondary peak, due to the
coarse grid-sampled properties of the synthetic SED library). We plan
to study the efficiency of the SED fitting algorithm in the estimate
of SFRs in the range $0.1 M_\odot/yr <$ SFR $< 10 M_\odot/yr$ as a
part of a larger programme, comparing the global performances of our
algorithm in recovering the physical properties of model galaxies with
complex star formation histories, starting from synthetic photometry
extracted from SAM simulated catalogues. A similar procedure has been
recently employed by \citet{Lee09} to a sample of high-redshift model
galaxies: they found that SED fitting underestimates SFR in their
sample and attributed this behaviour to to the different star
formation histories predicted by SAMs with respect to those assumed in
the reference synthetic SED libraries.

\begin{figure}
  \centering \includegraphics[width=7.25cm]{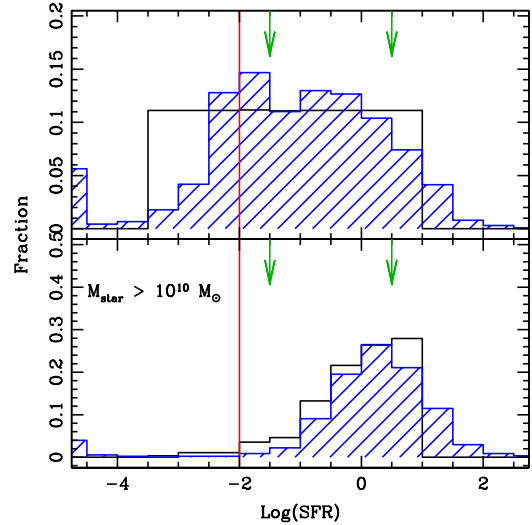}
\caption{Test on the SED fitting algorithm. {\it Upper panel}:
  comparison of the SFR distribution in a sample of $0.4<z<1.8$
  \morgana~model galaxies (solid black histogram), with the
  corresponding distribution of reconstructed SFRs, obtained applying
  the SED fitting algorithm to synthetic photometry (blue hatched
  histogram). Red line represents the limiting SFR considered in the
  \citet{Santini09} paper, while green arrow bracket the region
  containing the main features of the observed SFRF. {\it Lower
    Panel}: same as upper panel, but for a sample of $M_\star >
  10^{10} M_\odot$ model galaxies.
  \label{fig:fit_test}}
\end{figure}

As a preliminary test we consider a small sample of 5000 model
galaxies with $0.4<z<1.8$ taken from the same \morgana~realisation and
uniformly distributed in SFR. For the purpose of this test we do not
limit the sample to $M_\star > 10^{10} M_\odot$ model galaxies. For
each object, we compute a synthetic SED using the Radiative-Transfer
code {\sc grasil} (\citealt{Silva98}, see also
\citealt{Fontanot07b}). We then apply the SED fitting algorithm to the
corresponding synthetic photometry. In fig.~\ref{fig:fit_test} (upper
panel) we compare the original, flat, distribution of SFR in our
sample (black histogram) with the distribution of reconstructed SFRs
(blue shaded histograms). This test suggests that some depletion
effects in SFR distribution may be effective at SFR below our fiducial
limit (SFR $\sim 0.01 M_\odot/yr$); however, the SFR distribution in
the region corresponding to the secondary peak and the valley (roughly
identified by the green arrows) does not seem to be affected by any
obvious systematic effect. We also repeat the same analysis on a
subsample of $M_\star > 10^{10} M_\odot$ model galaxies, (lower
panel), whose distribution is more similar to the theoretical
SFRF. Again we obtain no obvious systematic effect in SED fitting
reconstruction. We also check that similar results hold if we split
the sample in smaller redshift bins, but with less statistical
significance, due to the reduced dimensions of the subsamples.

In fig.~\ref{fig:sfrf} we also show the predictions of our three SAMs
as coloured lines: the level of agreement between the 3 models is
striking, given the very different star formation and stellar feedback
schemes adopted. In particular we stress that the predicted redshift
evolution of the typical SFR$^\star$ is remarkably similar between all
models. This is due to the external constraints (i.e. the evolution of
the cosmic SFR) used to calibrate the models. The strongest
discrepancies are seen at the low-SFR end of the SFRF and clearly show
the importance of using constraints based on the SFRF, in order to
disentangle between different theoretical approaches to star formation
and stellar feedback.

The comparison between theoretical predictions and the observational
results from the GOODS-MUSIC survey shows a remarkably good agreement
at the high-SFR end, where the available $24 \mu$m observations are
deep enough to constrain its shape and evolution out to $z \sim 1.8$
(yellow shaded region in fig.~\ref{fig:sfrf}). Therefore we conclude
that SAMs provide a satisfactory description for the space density of
galaxies with SFR$>10 M_\odot/yr$ at moderate redshifts. It worth
stressing that accounting for observational errors is essential for
the correct interpretation of the comparison between model and data:
the error-free intrinsic SFRFs would underpredict the space density of
the strongest starbursts by a factor of a few.

\begin{figure*}
  \centering
  \includegraphics[width=10.5cm]{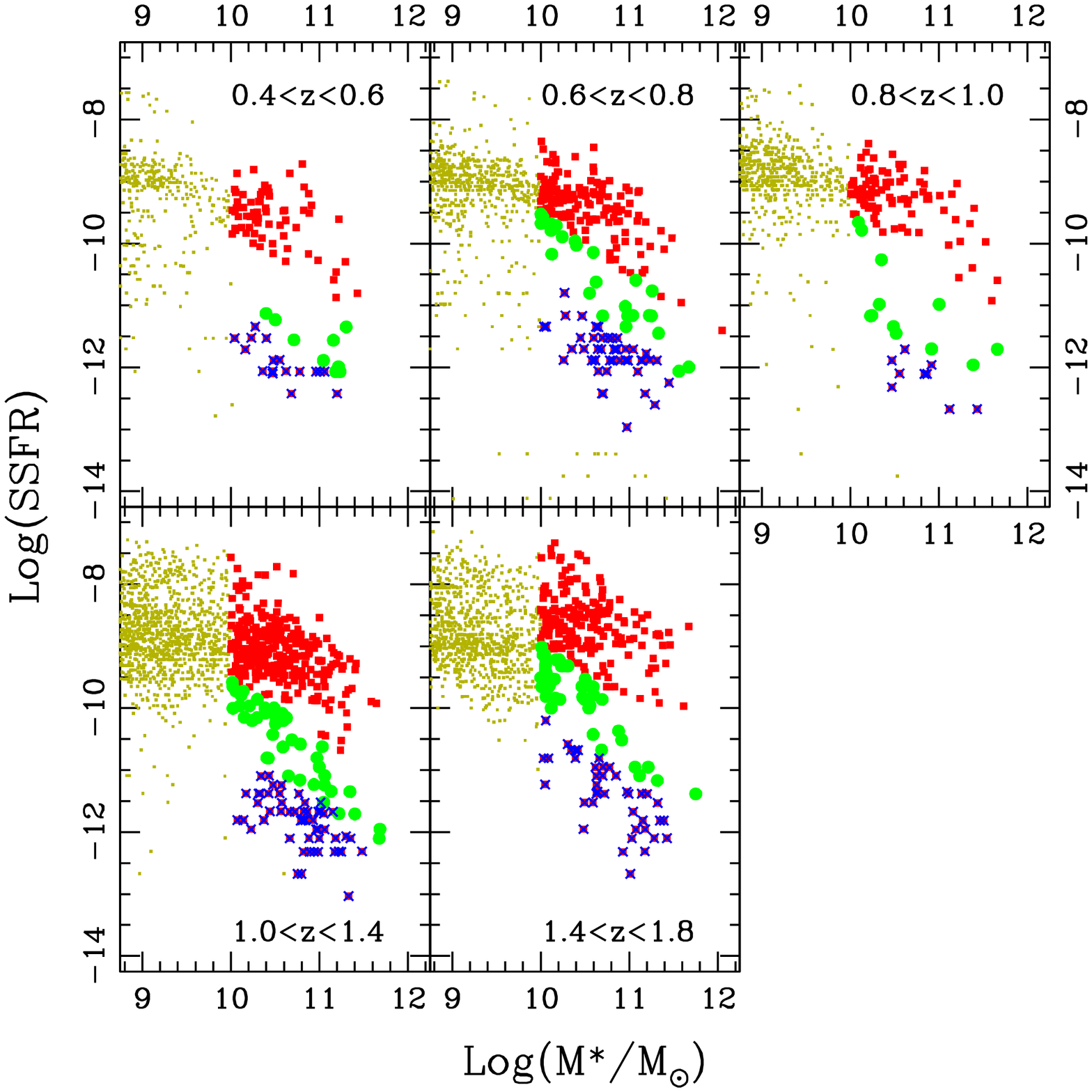}
  \includegraphics[width=10.5cm]{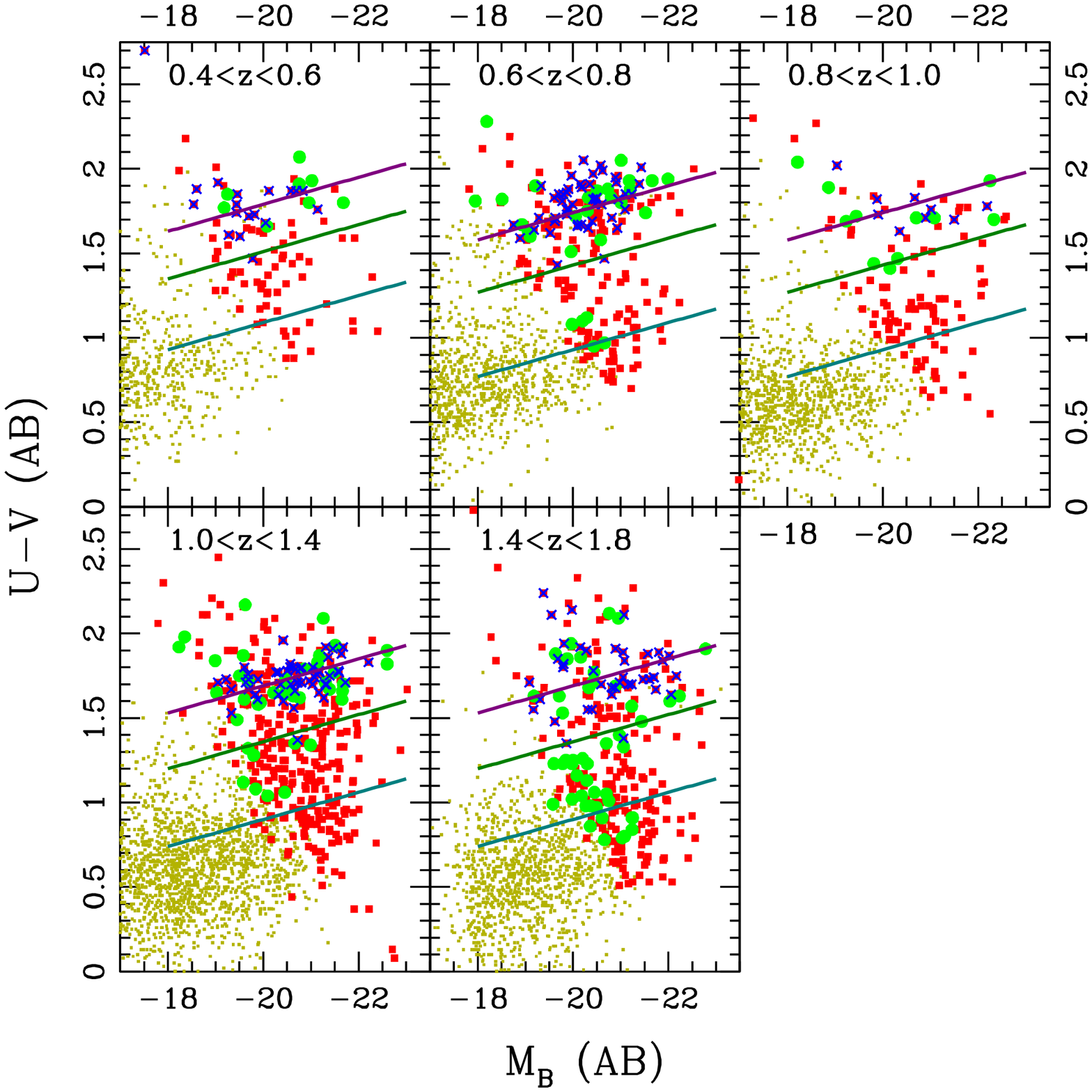}
\caption{{\it Upper panel:} Specific SFR versus $M_\star$ {\it Lower
    panel:} Rest-Frame colour-magnitude diagrams. In each panel the
  yellow dots refer to the $K$-limited GOODS-MUSIC sample; red dots
  mark sources with $M_\star > 10^{10} M_\odot$ and SFR $> 0.01
  M_\odot/yr$; blue crosses refer to galaxies belonging the secondary
  peak on the SFRF at each redshift bin, while green circles refer to
  galaxies with SFR intermediate between the two peaks. In the lower
  panel, dark red and blue lines refer to the mean $U-V$ colours along
  the red sequence and the blue cloud, while the dark green solid line
  reproduces the divider between the blue and red populations (see
  \citealt{Salimbeni08} for more details).
  \label{fig:colour_dia}}
\end{figure*}

Observations at $24 \mu$m are also able to constrain the position of
the peak of the SFRF out to $z \sim 1$ (and give some indications for
its position at higher redshifts). In order to get more insight on the
evolution of the typical SFR of galaxies at different redshifts, we
repeat on the data the same fitting procedure described in
sec.~\ref{sec:models}. Given the double peaked shape of the observed
SFRF both a double power law and a Schechter function are no longer a
good representation of the SFRF over the whole SFR range. Therefore,
we chose to repeat the fitting procedure both on the reference sample
($SFR>0.01 M_\odot/yr$), and on a smaller sample containing only
sources with $SFR>1 M_\odot/yr$ (i.e. focusing on the primary SFRF
peak). This choice has no major impact on our conclusions; as expected
only the low-SFR-end slope $\alpha$ show a relevant variation between
the two samples. We collect our final results in
tab.~\ref{tab:bf_obs}. The results show that the SFRF, as depicted by
the GOODS-MUSIC survey, shows a remarkable density evolution (PDE) in
the $0.8<z<1.8$ range, and a milder SFR$^\star$ evolution (PSE) over
the whole redshift range. The theoretical SFRF is able to reproduce
the SFR$^\star$ evolution, although it peaks systematically at lower
typical SFR than the observational constraints, but its density
evolution is negligible. This represents a first relevant discrepancy
between models and data and it is due to the well documented
overprediction of the space density of $10^{10} M_\odot < M_\star <
10^{11} M_\odot$ galaxies at $1.0 < z < 1.8$ seen in the models and in
the data; \citet{Fontanot09b} showed that SAMs predict for these
galaxies space densities very similar to the local value out to
$z\sim1.8$, while their observed stellar mass function show a
remarkable evolution (their fig.1). This implies that the negligible
density evolution predicted for the SFRF at $z>1.0$ is related to the
too efficient formation of intermediate mass galaxies in SAMs (see
also \citealt{LoFaro09}).
\begin{table*}
  \caption{Best fit parameters for the evolution of the SFR function
    in the GOODS-MUSIC Catalogue.}
  \label{tab:bf_obs}
  \renewcommand{\footnoterule}{}
  \centering
  \begin{tabular}{cccccccc}
    \hline
    Model & $\Phi^{\star} (\times 10^{-2})$ & $SFR^{\star}$ & $k_{\rm SFR}$ & $k_{\Phi}$ & $\alpha$ & $\beta$ & $\chi^2$\\
    \hline
    \multicolumn{8}{c}{$Log(SFR)>-2.0$} \\
    \hline
    PDE     & $0.37 \pm 0.11$ & $1.13 \pm 0.13$ & $0.70 \pm 0.58$ & --- & unconstr. & $-1.77 \pm 0.30$ & $2.7$ \\
    PSE     & $0.74 \pm 0.20$ & $1.30 \pm 0.07$ & --- & $-1.09 \pm 0.60$ & unconstr. & $-1.85 \pm 0.29$ & $1.6$ \\
    PDE+PSE & $0.88 \pm 0.16$ & $0.81 \pm 0.11$ & $2.07 \pm 0.44$ & $-1.22 \pm 0.22$ & unconstr. & $-1.43 \pm 0.31$ & $1.2$ \\
    \hline
    \multicolumn{8}{c}{$Log(SFR)>0.0$} \\
    \hline
    PDE     & $0.30 \pm 0.05$ & $1.02 \pm 0.19$ & $0.93 \pm 0.35$ & --- & unconstr. & $-1.84 \pm 0.21$ & $1.5$ \\
    PSE     & $0.58 \pm 0.16$ & $1.31 \pm 0.11$ & --- & $-0.72 \pm 0.23$ & unconstr. & $-1.48 \pm 0.24$ & $0.9$ \\
    PDE+PSE & $0.55 \pm 0.05$ & $0.54 \pm 0.15$ & $2.09 \pm 0.38$ & $-0.56 \pm 0.11$ & unconstr. & $-1.43 \pm 0.24$ & $0.6$ \\
    \hline
  \end{tabular}
\end{table*}

The most interesting discrepancies between models and data are indeed
seen at the low-SFR end: all SAMs predict a featureless low-SFR end,
whose slope (and its redshift evolution) depends on the details of the
treatment of SFR and stellar feedback. The most relevant difference
between data and model prediction is the fact that all models
overpredict the abundance of SFR$\sim1 M_\odot/yr$ sources. It is also
interesting to notice that no model reproduces the double peaked
feature of the observed SFRF.

There are at least two possible interpretations of our results. First
(scenario A), as it has been suggested by many authors (see e.g.,
\citealt{DeLucia07b, Wang07}), SAMs predict too low SFRs for most of
the galaxies in the mass range of our interest: they have SFRs well
below the detection limit of our sample, and, overall, they are too
passive at intermediate redshifts \citep{Fontanot09b}. The three
models indeed predict a secondary peak in their SFR distribution, and
it consists of completely passive galaxies (SFR$=0$, see e.g. figure 4
in \citealt{Fontanot09b}). In this scenario, we could interpret the
observed secondary peak (at low, but still detectable SFRs), if
confirmed by deeper observations, as an evidence that models are not
able to reproduce the correct space density for galaxies migrating
from the blue cloud to the red sequence (see also
\citealt{Cassata07}), and we could interpret it as an indication that
galaxies in the GOODS-MUSIC sample move in the colour-magnitude
diagram at a slower pace with respect to theoretical
expectations. This scenario is also consistent with the recent
findings of \citet{Lagos11}: they revisited the definition of the star
formation law in the context of the {\sc galform} SAM \citep{Baugh05,
  Bower06} taking into account recent results on the relation between
star formation and the atomic and molecular gas content of galaxies
(see e.g.  \citealt{BlitzRosolowsky06, Krumholz09}). They showed that
the properties of galaxies with low SFRs (and their redshift evolution
on the $M_\star$ versus SFR plane) are particularly sensitive to the
details of the assumed star formation law, and suitable modifications
to the standard prescriptions may indeed lead to bimodal SFR
distributions in the direction needed to explain our results.

An alternative scenario (scenario B) requires small SFR retriggering
events in already red galaxies (see e.g. \citealt{Treu05}): these
events might be linked to the dynamical evolution of galaxies in dense
environments (mergers, interactions). Therefore this also implies a
dependence of these secondary SFR events on the environment, which in
principle could be accessible by larger surveys (see
e.g. \citealt{Cucciati10} and references herein). Another possible
origin for small and late star formation episodes in massive galaxies
might be related to the duty cycles associated to the ``radio-mode''
AGN activity \citep{Fontanot10}: if the injection of energy from the
central Black Hole into the surrounding Dark Matter halo is not a
continuous event, some cold gas might still be allowed to infall onto
massive galaxies.

In order to get more insight in the properties of galaxies responsible
for this feature, we consider the distribution of the GOODS-MUSIC
galaxies both in the specific SFR versus $M_\star$ and in rest-frame
colour-magnitude diagram (as reconstructed from the SED fitting
algorithm) in fig~\ref{fig:colour_dia}. In both panels, we mark the
$M_\star>10^{10} M_\odot$ galaxies as red squares; blue crosses and
green circles refer to the galaxies belonging to the secondary peak
and lying between the two peaks respectively. Dark red, green and blue
lines in the lower panel reproduce the locus of red sequence, green
valley and blue cloud as defined in \citet{Salimbeni08}. 

We expect the two scenarios we discuss in the previous paragraph to
predict a different colour evolution for galaxies. In scenario A, we
assume a continuity between the colour of galaxies belonging to the
primary peak, valley and secondary peak, in terms of blue cloud,
``green valley'' and red sequence. Since scenario A assumes a
continuous motion of galaxies from the blue cloud to the red sequence,
we expect galaxies with different SFRs to populate well defined
regions of the colour-magnitude diagram. The relative abundance of
galaxies in the different regions of the diagram is an indication of
the time they spend in the different phases.

On the other hand, in scenario B we expect a larger spread of galaxies
belonging to the secondary peak in the colour magnitude diagram. The
physical processes responsible for the retriggering have a strong
dependence on the stellar mass and/or environment of galaxies, and are
almost insensitive to the position of the galaxy in the
colour-magnitude diagram at the time of the retriggering. Therefore,
the relative tightness of the locus where secondary peak objects are
found in the rest-frame colour-magnitude diagrams is in qualitative
agreement with the former scenario. On the other hand, galaxies with
SFR intermediate between the two peaks show a much wider distribution
in fig.~\ref{fig:colour_dia}. We warn the reader that these are
qualitative conclusions that rely on the assumption that the $U-V$
colour is a fair tracer of the SFR activity, and that this colour is
also sensitive enough to the low-levels of SFR for galaxies in the
secondary peak. The fact that most of SFR estimates for sources below
the SFRF peak have been derived from SED fitting also prevents firm
conclusions on this issue from GOODS-MUSIC data alone. A more
quantitative analysis will require better estimates for SFR levels in
galaxy population and better tracers of physical state of galaxies
belonging to the different regions of the colour-magnitude diagram,
and it is beyond the aims of the present work.

\section{Conclusions}\label{sec:summary}

In this paper we consider the GOODS-MUSIC sample to study the SFR
function and its redshift evolution in the redshift range $0.4<z<1.8$,
just after the peak in the cosmic SFR, a fundamental epoch to
understand the physical processes responsible for the decline of star
forming activity as a function of galaxy stellar mass. In order to
reduce the effect of completeness corrections, we restrict our
analysis to galaxies with $M_\star > 10^{10} M_\odot$, since at these
stellar masses the GOODS-MUSIC is expected to be a complete sample. We
thus compute the SFRF using SFR estimates coming from both SED fitting
and/or a combination of UV light with $24 \mu$m observations. We
compare the resulting SFRF with the predictions of 3 independent
semi-analytical models of galaxy formation and evolution
(\citealt{Wang08,Somerville08} and \morgana) and we obtain the
following results:

\begin{itemize}
\item{The agreement between the observed and predicted high-SFR end
  (SFR$>10 M_\odot/yr$) of the SFRF is good over the whole redshift
  range (once the observational errors are taken into account). This
  result implies that theoretical models are able to reproduce the
  space density evolution of the strongest starbursts. It is worth
  noting that the predictions of the three models are remarkably
  similar in this SFR range.}
\item{The SFRF of massive galaxies, as seen by the GOODS-MUSIC is
  characterised by a complex evolution both in the typical SFR and in
  the normalisation. Theoretical models are able to cope with the
  SFR$^\star$ evolution, but they predict a negligible density
  evolution. In particular, all models overpredict the space density
  of SFR $ \sim 1 M_\odot/yr$ galaxies. We interpret this behaviour as
  due to the well documented excess of intermediate mass galaxies
  ($10^{10} M_\odot < M_\star < 10^{11} M_\odot$) in SAMs
  \citep{Fontanot09b}.}
\item{The observed low-SFR end of SFRF is characterised by a double
  peaked shape. Despite the uncertainties in the determination of such
  low SFRs from SED fitting methods, it is worth noting that none of
  the theoretical model we consider is able to reproduce this feature,
  and all of them predict a smooth decrease of the space density of
  low-SFR galaxies, with a well defined power-law slope
  $\alpha$. Since this is also the SFR range where models differ most,
  each of them predicting a different value for $\alpha$, we stress
  that future observations providing stronger constraints on the
  low-SFR end of the SFRF will be of fundamental importance to
  understand the physical mechanisms responsible for the decline of
  SFR since $z\sim2$ in massive galaxies.}
\end{itemize}

The main result of this paper lies in the analysis at the high-SFR
end, since this region is completely sampled by $24 \mu$m
observations, which are critical for the recovery of the SFR
levels. Also the evolution of the typical SFR$^\star$ is well sampled
by means of IR data out to $z \sim 1.0$, and allow us to put strong
constraint on this relevant discrepancy seen between data and models.
Unfortunately, for lower SFR levels this information is not currently
available, and we have to rely entirely on the results of SED
fitting. Our tests suggest that the use of SED fitting results does
not introduce any obvious distortion in the SFR distribution at
intermediate SFR values, but we could not completely exclude the
double peaked feature in the SFRF to be due to systematics in the SED
fitting algorithm. From optical photometry alone is not possible to
uniquely associate SFR $ \sim 1 M_\odot/yr$ galaxies (the region were
models and data differ most and roughly corresponding to the region
between the two peaks of the SFRF) with known populations in the
colour-magnitude diagram, i.e. the green valley or the blue cloud. The
red $U-V$ colours, with a relatively small scatter, of galaxies
belonging to the putative secondary peak provide only a qualitative
support to a scenario where star formation in massive galaxies
decrease at a slower pace with respect to theoretical expectations. If
confirmed, this double peaked shape of the SFRF would thus represent a
critical discrepancy between models and observations, since all SAMs
we consider predict a smooth, featureless, single slope power-law
shape for the SFRF. Modifications in the assumed star formation law,
taking into account the relation between star formation and the atomic
and molecular gas content of galaxies may alleviate this tension,
since the properties of model galaxies with low SFR levels are very
sensitive to this modeling layer \citep[see e.g.]{Lagos11}.

On the other hand, thanks to the new facilities (e.g. {\it Herschel}),
future observations in the IR region, able to characterise the
properties of the IR peak, will finally determine the SFR levels of
these objects and the relevance of this feature. This is a fundamental
task, since the low-SFR end of the SFRF is the region where SAM
predictions differ most and therefore provides potentially strong
constraints to the different approaches to star formation and stellar
feedback.

\section*{Acknowledgements}
We thank Gabriella De Lucia and Pierluigi Monaco for stimulating
discussions. We are grateful to Jie Wang for letting us use the
outputs of his simulations. FF acknowledges the support of an
INAF-OATs fellowship granted on 'Basic Research' funds and financial
contribution from the ASI project ``IR spectroscopy of the Highest
Redshift BH candidates'' (agreement ASI-INAF 1/009/10/00).

\bibliographystyle{mn2e} 
\bibliography{fontanot}

\end{document}